\documentclass[epj]{svjour}
\usepackage{graphicx}        

\usepackage{amsmath}
\usepackage{amssymb}

\newcommand \beq{\begin{eqnarray}}
\newcommand \eeq{\end{eqnarray}}

\begin{document}
\title{Photon and dilepton production in the Quark-Gluon plasma:\\
  perturbation theory vs lattice QCD}
\author{Jean-Paul Blaizot\inst{1}\and
Fran\c cois Gelis\inst{2}}

%
%
\institute{ECT*, Villa Tambosi, Strada delle Tabarelle 286, 38050
Villazzano (TN) Italy \and Service de Physique Th\'eorique, CEA
Saclay, 91191 Gif-sur-Yvette cedex, France}
%
%
\abstract{This talk reviews the status of QCD calculations of photon
    and dilepton production rates in a Quark-Gluon plasma. Theses rates
    are known to order ${\cal O}(\alpha_s)$. Their calculations involve
    various resummations to account for well identified physical effects
    that are briefly described. Lattice calculations of the spectral
    functions give also access to the dilepton rates.  Comparison with
    perturbative results points to inconsistencies in both approaches
    when the dilepton energy becomes small.\\
    \\
    {\bf Preprint no:~}ECT-05-04, SPhT-T05/054
\PACS{  {12.38.Mh}{Quark-gluon plasma} \and
        {13.85.Qk}{Electromagnetic probes}
       } 
} 

\date{}
\titlerunning{Photon and dilepton production in the QGP}
\authorrunning{Jean-Paul Blaizot\and Fran\c cois Gelis}
\maketitle

\section{Introduction}
\label{sec:1} Photons or lepton pairs are produced at  various
stages of a nucleus-nucleus collision. Prompt photons and large mass
dileptons are produced in the initial partonic collisions. Their rates
can be calculated using zero temperature perturbative QCD. They
populate the high energy part of the spectrum.  All other photons or
dileptons result from secondary interactions between the produced
particles. We focus here on the photons which are produced in a
thermalized quark-gluon plasma (QGP). Their rates can be calculated
using equilibrium thermal field theory. We shall not discuss how the
rates can be combined with the space time evolution of nucleus-nucleus
collisions in order to obtain the observed yields
\cite{Sriva1,SrivaS1,HuoviRR1,AlamSRHS1}).  Nor shall we discuss
photons produced in the hadronic phase (see the talk by Charles Gale
in these proceedings). This talk builds on Ref.~\cite{GELIS-Quark
Matter} and extends some of the discussions presented there.

The photon production rate can be expressed in terms of the the
current-current correlator $\langle j_\mu(0)j_\nu(x)\rangle$, where
the electromagnetic current is $j_\mu(x)=\overline{\psi}(x)\gamma_\mu
\psi(x)$. To leading order in the electromagnetic fine structure
constant $\alpha$, the photon production rate reads
\cite{Weldo3,GaleK1} ($\omega^2={\boldsymbol q}^2$):
\begin{eqnarray} { {\omega\frac{dN_\gamma}{d^4 x\, d^3{q}}}}&=
&-\frac{e^2 g^{\mu\nu}}{2(2\pi)^3} \Pi^<_{\mu\nu}(\omega,{
q})\nonumber\\ &=& \frac{e^2}{(2\pi)^3}
\frac{g_{\mu\nu}}{e^{\omega/T}-1} \;{ {\rm Im}\,\Pi_{\rm
ret}^{\mu\nu}(\omega,{ q})}\; .
\label{eq:rate-1}
\end{eqnarray}
where $ \Pi^<_{\mu\nu}(\omega,{ q})$ is the electromagnetic
polarization tensor:
\begin{eqnarray}
    \Pi^<_{\mu\nu}(\omega,{ q})=\int{\rm d}^4x\,{\rm e}^{iQ\cdot
x}\,\langle j_\mu(0)j_\nu(x)\rangle\; .
    \end{eqnarray}
The second of eqs.~(\ref{eq:rate-1}) gives the photon production rate
in terms of the retarded polarization tensor. A similar formula
exists for lepton pairs ($Q=(\omega,{\boldsymbol q}), \,Q^2\equiv
\omega^2-{\boldsymbol q}^2>0$):
\begin{eqnarray}
      { {\frac{dN_{l^+l^-}}{d^4 x d^4{Q}}}}= \frac{e^4}{3(2\pi)^4 Q^2}
   \frac{Bg_{\mu\nu}}{e^{\omega/T}-1} \;{ {\rm Im}\,\Pi_{\rm
   ret}^{\mu\nu}(\omega,{ q})}\; ,
   \label{eq:rate-2}
\end{eqnarray}
where the phase space factor
\begin{eqnarray}\label{threshold}
B\equiv\left( 1+\frac{2m_l^2}{Q^2}\right)\left(
1-\frac{4m_l^2}{Q^2}\right)^{1/2}
\end{eqnarray}
indicates a threshold at $Q^2=4m_l^2$, with $m_l$ the mass of the lepton.

In the first part of this talk, we shall review the analytical
calculations of the rate, based on weak coupling techniques. Then we
shall briefly discuss the estimates obtained from lattice
determinations of spectral functions.

\section{Weak coupling  calculations}
\subsection{Leading order}
\begin{figure}[htbp]\centering
\resizebox*{!}{1.1cm}{\includegraphics{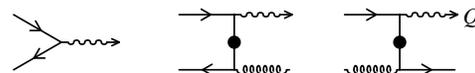}}
\caption{\label{fig:processes-1} Real processes contributing to photon
    and dilepton production up to ${\cal O}(\alpha_s)$.  One loop
    virtual corrections to the first process also contribute at this
    order.}
\end{figure}
The leading order contribution to the dilepton rate is obtained from
the one-loop contribution to the polarization tensor, and corresponds
to the Drell-Yan process illustrated by the diagram on the left of
fig.~\ref{fig:processes-1} (only the production of the virtual
photon is represented). It was evaluated for a QGP in \cite{McLerT1}.

\subsection{First perturbative corrections}
The corrections of order ${\cal O}(\alpha_s)$, where
$\alpha_s\equiv g^2/4\pi$, with $g$ the QCD gauge coupling, correspond to
the two diagrams in the right of figure \ref{fig:processes-1}. Their
calculation reveals two problems. For the dilepton rate ($Q^2>0$) in a
plasma of massless quarks and gluons, each individual contribution to
eq.~(\ref{eq:rate-2}) contains a mass singularity, and it is only
after a careful summation of all the real and virtual corrections that
one gets a finite result \cite{BaierPS2,AltheAB1,AltheR1}.  This is
nothing but a manifestation of the KLN theorem
\cite{LeeN1,Kinoshita}. In the case of real photons ($Q^2\to 0^+$) a
new singularity appears, with contributions of the form
\begin{eqnarray}
\label{eqn3}{\rm
    Im}\,\Pi_{\rm ret}(\omega,{\boldsymbol q}) \propto \alpha\alpha_s
\ln(\omega T/Q^2)
\end{eqnarray}
at small $Q^2$. The singularity
originates from the presence of intermediate massless quarks (the
``vertical'' propagators in the two diagrams in the right of
fig.~\ref{fig:processes-1}). As we shall see, plasma effects
induce effective masses and cure part of the difficulty.

\subsection{Scales and degrees of freedom in a quark-gluon plasma}
At this point it is useful to recall some basic properties of a
quark-gluon plasma in the weak coupling regime, i.e. at sufficiently
high temperature \cite{Blaizot:2001nr}. This regime is characterized
by a hierarchy of momentum scales.  Most of the plasma particles have
momenta of the order of the temperature $T$, and since their density
is of order $T^3$, $T$ is also the inverse of the inter-particle
distance. Besides, collective excitations can develop in the
system. Such collective phenomena are particularly important at the
scale $gT$, where $g$ is the gauge coupling (the reason why these
excitations are called collective is that, when $g\ll 1$, their
wavelength $\sim 1/gT$ is large compared to the inter-particle distance
$\sim 1/T$, so that many particles participate in the excitation).
Systematic corrections to the propagation and interactions of such
collective excitations involve the resummation of the so-called ``hard
thermal loops" \cite{BraatP1,FrenkT1}. Note that the soft collective
modes also modify the spectrum of hard particles, giving them a mass
that we shall refer to as $m_\infty$ ($\sim gT$). Finally, another
scale plays an important role in a quark-gluon plasma: this is the
scale $g^2T$ where perturbation theory breaks down because of the
presence of unscreened magnetic fluctuations. The scale $g^2T$
characterizes also the rate of collisions with small ($\sim gT$)
momentum transfer. To see that write the scattering cross section as
$\sigma=\int{\rm d}q^2({\rm d}\sigma/{\rm d}q^2)$, where typically $
{\rm d}\sigma/{\rm
d}q^2\sim g^4/q^4$. The collision rate is
$\gamma=n\sigma$, so that, with $ n\sim T^3$, $ \gamma\sim
g^4\,T^3\,\int {\rm d}q^2 /{q^4}.  $ The infrared divergence of the
integral is cut-off by the screening mass $m_D$ ($\sim gT$) ($m_D$ is
an example of a ``hard thermal loop'') leaving a finite result $
\gamma\sim g^4T^3/{m_D^2}\sim g^2T.  $ This simple estimate applies
when collisions involve dominantly small momentum transfer $\sim
gT$. However, when calculating the effect of collisions on transport
properties, the dominant collisions involve large angle scattering and
the infrared cut-off is actually taking place at a larger momentum
scale, of order $T$. Thus, most transport coefficients end up being of
order $g^4T\ln(1/g)$ \cite{Baym:1990uj}.

\subsection{Resummation of hard thermal loops}
We can now return to eq.~(\ref{eqn3}) and observe that the logarithmic
singularity at $Q^2\to 0$ is due to the exchange of a soft massless
quark. Once the HTL correction is included on the quark propagator,
the quark effectively acquires a mass $m_{\infty}$ of order
$gT$ ( $m_{\infty}^2=\pi\alpha_s C_f T^2$ with
$C_f\equiv(N_c^2-1)/2N_c$). By taking into account this thermal
correction one obtains a finite photon polarization tensor
\cite{KapusLS1,BaierNNR1}. For hard photons, it reads:
\begin{eqnarray}
{\rm Im}\,\Pi_{\rm ret}{}^\mu{}_\mu(\omega,{\boldsymbol q})=4\pi
\frac{5\alpha\alpha_s}{9}T^2 \left[\ln\left(\frac{\omega
T}{m_{\infty}^2}\right)
+{\rm const}\right]\; .
\end{eqnarray}
The numerical factor $5/9$ is the sum of the quark electric charges
squared for $2$ flavors (u and d); for $3$ flavors (u, d and s), this
factor should be replaced by $6/9$.  This formula indicates how the
infrared problem is cured: $Q^2$ in Eq.~(\ref{eqn3}) is effectively
replaced by $m_{\infty}^2$ in the logarithm as soon as $Q^2$ becomes
small compared to $m_{\infty}^2$.

    \begin{figure}[htbp]
\centering \resizebox*{!}{1.3cm}{\includegraphics{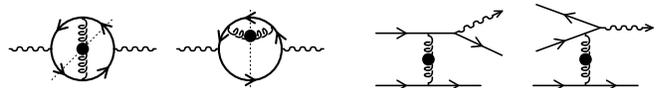}}
\caption{\label{fig:processes-2} Processes that are
    promoted to ${\cal O}(\alpha_s)$ by collinear singularities.}
\end{figure}

This, however, is not the final answer for the photon and dilepton
rates at ${\cal O}(\alpha_s)$. Indeed there are other processes,
formally of higher order, which are strongly enhanced by collinear
singularities and become effectively of order $\alpha_s$. This was
first realized for soft photon production by quark bremsstrahlung
\cite{AurenGKP1,AurenGKP2} (third diagram of
fig.~\ref{fig:processes-2}, starting from the left). The diagram on
the right of figure \ref{fig:processes-2} shares the same property,
but contributes significantly only to hard photon production
\cite{AurenGKZ1}, due to phase-space suppression in the case of soft
photons. Note that a naive power counting would indicate that these
two diagrams contribute to ${\cal O}(\alpha_s^2)$.  These two diagrams
represent collision processes: they originate from cutting a loop
insertion in the gluon propagator in order to get the imaginary part
(the cuts are indicated by the dotted lines in the two diagrams on the
left of fig.~\ref{fig:processes-2}).

\begin{figure}[htbp]\centering
\resizebox*{!}{1.5cm}{\includegraphics{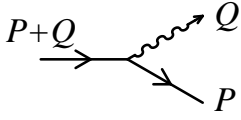}}
\caption{\label{fig:virtuality} The virtual quark of momentum $P+Q$
emitting a real photon ($Q^2=0$) and an on-shell quark of momentum $P$
($P^2=m^2$).}
\end{figure}

In order to understand the origin of the ``collinear enhancement'',
let us focus on the quark propagator between the quark-gluon vertex
and the photon emission. The virtuality of this off-shell quark is
easily estimated (see fig.~\ref{fig:virtuality}):
\begin{eqnarray}
   (P+Q)^2-m_\infty^2=2P\cdot Q\approx \frac{m_\perp^2}{p_z}\,\omega\; ,
\end{eqnarray}
where $ m_\perp^2\equiv p_\perp^2+m_\infty^2$. Thus the virtuality of
this quark can become very small if the quark is massless and the
photon is emitted forward ( $p_\perp\to 0$).  Of course, the quark
thermal mass $m_{\infty}$ prevents these diagrams from being truly
singular. However, contrary to the ${\cal O}(\alpha_s)$ diagrams, the
singularity is here linear instead of logarithmic, and brings a factor
$T^2/m_{\infty}^2\sim 1/\alpha_s$. Combined with the $\alpha_s^2$ that
comes from the vertices, the singularity turns the contribution of
these diagrams into an order ${\cal O}(\alpha_s)$ contribution. This
was evaluated in \cite{AurenGKP2} and \cite{AurenGKZ1} (see also
\cite{Mohan2} and \cite{SteffT1} where an erroneous factor $4$ was
pointed out), and a closed expression was obtained in
\cite{AurenGZ4,AurenGZ3}. The result is of the form:
   \begin{eqnarray}\label{eqn6}
   {\rm Im}\,\Pi_{\rm ret}{}^\mu{}_\mu(\omega,{\boldsymbol q})= {\rm const\ }
   \alpha\alpha_s \left[\pi^2 \frac{T^3}{\omega}+\omega T\right]\; .
   \end{eqnarray}
In this formula, the term in $1/\omega$ dominates for soft photons
and comes from the brem\-sstra\-hlung diagram, while the term in
$\omega$ comes from the second diagram and  dominates the rate of very hard
photons ($\omega \gg T$).

It is worth mentioning that the purely numerical prefactor (not
written explicitly in the previous formula) is a function of the ratio
of the quark thermal mass $m_{\infty}$ to the gluon Debye mass
$m_{_D}$.  In the HTL approximation, this ratio is a constant
independent of the coupling and temperature, that depends only on the
number of colors and flavors; for $3$ colors and $N_f$ flavors, this
ratio is $m_{\infty}/m_{_D}=\sqrt{2/(6+N_f)}$.

This enhancement due to a quasi-collinear emission of the photon also
occurs for the emission of virtual photons with a small invariant mass
($Q^2\equiv \omega^2-{\boldsymbol q}^2 \ll \omega^2$), but becomes
less and less important when the photon invariant mass increases. For
virtual photons with vanishing momentum (i.e. for which the invariant 
mass is maximal, at a given
energy), the two-loop diagrams contribute only at the order $g^3$
\cite{Aurenche:2003ac}, instead of $g^2$ for real photons.

\subsection{LPM resummation}
The collinear enhancement that we have identified on so\-me of the
order $\alpha_s^2$ processes affects in fact an infinite set of
processes.  In order to explain the issue in physical terms, it is
convenient to introduce the concept of {\sl photon formation time}.
Let us return to the process in figure~\ref{fig:virtuality}.  The
photon formation time can be identified with the lifetime of the
virtual quark, which is itself related to its virtuality by the
uncertainty principle. A simple calculation gives:
\begin{eqnarray}\label{dEform}
   \delta E= q+E_p-E_{p+q}\approx
   \frac{m_\perp^2}{2}\frac{\omega}{p_z(p_z+\omega)}\; ,
\end{eqnarray}
where the 3-momentum of the photon defines the longitudinal axis. The
formation time is $t_{_F} = 1/\delta E$.  The collinear enhancement in
the diagrams of figure \ref{fig:processes-2}, due to the small
virtuality of the quark that emits the photon, can be rephrased in
terms of the large photon formation time. Typically, in a quark-gluon
plasma, we have $m_\perp^2\sim g^2 T^2$, while $p_z\sim T$, so that
$\delta E$ in eq.~(\ref{dEform}) is $\delta E\sim g^2 T$ for
$\omega\sim T$. That is, the photon formation time is of the same
order of magnitude or larger than the quark mean free path between two
soft collisions, i.e. $t_{_F}\sim 1/\gamma$, where $\gamma\sim g^2 T$
is the collision rate estimated earlier. Note that the estimate done
earlier is indeed the relevant collision time scale for the production
of photons almost collinear with the charge particle, that is with a
typical transverse momentum of order $gT$: this is the kinematical
condition leading to the enhancement that we are discussing (by
``enhancement'', we mean, as earlier, the phenomenon by which higher
order diagrams turn out to contribute at the same order as a given
elementary process).  The sensitivity to the collisional width found
in \cite{AurenGZ2} occurs in the same kinematical conditions.  When
the mean free path becomes of the order of, or smaller than the photon
formation time, the effects of multiple collisions on the production
process can no longer be ignored. The result of such multiple
scattering is to reduce the rate compared to what it would be if all
collisions could be treated as independent source of photon
production. This phenomenon is known as the Landau Pomeranchuk Migdal
(LPM) effect \cite{LandaP1,LandaP2,Migda1}.
   \begin{figure}[ht]
\begin{center}
\resizebox*{5cm}{!}{\includegraphics{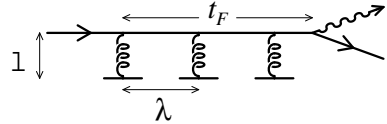}}
\end{center}
\caption{\label{fig:LPM} Diagram illustrating the conditions under
which multiple collisions need to be taken into account in the photon
or dilepton production process: $l$ is the typical range of the
interaction producing the collisions; $\lambda$ is the mean free path,
and $t_F$ the photon formation time.}
\end{figure}

\begin{figure}[ht]
\begin{center}
\resizebox*{5cm}{!}{\includegraphics{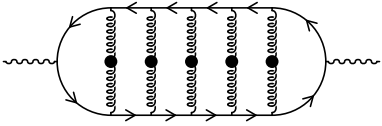}}
\end{center}
\caption{\label{fig:equations}  Resummation of ladder diagrams.}
\end{figure}

While the early treatment of the multiple scattering was done in terms
of kinetic equations, modern discussions used the language of quantum
field theory. The multiple scattering diagrams that must be resummed
are the ladder diagrams, as illustrated in figure~\ref{fig:equations}
(these are the typical diagrams that are taken into account by a
Boltzmann equation \cite{Blaizot:1999xk}). Cancellations between
self-energy corrections and vertex corrections remove any sensitivity
to the magnetic scale: physically, such cancellations reflect the fact
that ultrasoft scatterings (at momentum scale softer than $gT$) are
not efficient enough to produce a photon. A thorough diagrammatic
analysis explaining why it is the ladder family of diagrams that needs
to be resummed in order to obtain the complete leading ${\cal
O}(\alpha_s)$ photon rate is presented in \cite{ArnolMY1}.

   In the recent literature, the resummation of the ladder
   diagrams is presented as follows. The photon polarization
tensor is written explicitly as \cite{ArnolMY1,ArnolMY2,ArnolMY3}:
\begin{eqnarray}
{\rm Im}\,\Pi_{\rm ret}{}^\mu{}_\mu(Q)&\approx& {\alpha N_c}
\int_{-\infty}^{+\infty}\!\!dp_0 \,[n_{_{F}}(r_0)-n_{_{F}}(p_0)]\;
\frac{p_0^2+r_0^2}{(p_0r_0)^2} \nonumber\\  &\times& {\rm Re}\int
\frac{d^2{\boldsymbol p}_\perp}{(2\pi)^2}\; {\boldsymbol
p}_\perp\cdot{\boldsymbol f}({\boldsymbol p}_\perp)\; ,
\label{eq:AMY}
\end{eqnarray}
with $r_0\equiv p_0+\omega$, $n_{_{F}}(p_0)\equiv 1/(\exp(p_0/T)+1)$
the Fermi-Dirac statistical weight, and where the dimensionless
function ${\boldsymbol f}({\boldsymbol p}_\perp)$ denotes the resummed
vertex connecting the quark line and the transverse modes of the
photon\footnote{For the emission of real (massless) photons, only the
transverse polarizations of the photon matter.}. This function is
dotted into a bare vertex, which is proportional to ${\boldsymbol
p}_\perp$.  The equation that determines the value of ${\boldsymbol
f}({\boldsymbol p}_\perp)$ is a Bethe-Salpeter equation that resums
all the ladder corrections to the vertex
\cite{ArnolMY1,ArnolMY2,ArnolMY3}:
\begin{eqnarray}
\frac{i}{t_{_{F}}}{\boldsymbol f}({\boldsymbol p}_\perp) &=&
2{\boldsymbol p}_\perp \nonumber\\&\hspace{-1cm}+&\hspace{-0.5cm}4\pi
\alpha_s C_f T \!\! \int
\frac{d^2{\boldsymbol l}_\perp}{(2\pi)^2} \, {\cal C}({\boldsymbol
l}_\perp) \, [{\boldsymbol f}({\boldsymbol p}_\perp+{\boldsymbol
l}_\perp)-{\boldsymbol f}({\boldsymbol p}_\perp)]\; ,\nonumber\\
\label{eq:integ-f}
\end{eqnarray}
where $t_{_{F}}$ is the formation time, $t_{_{F}}=1/\delta E$, with
$\delta E$ given by eq.~(\ref{dEform}), and where the collision kernel
has the following expression  \cite{AurenGZ4}:
\begin{eqnarray}
{\cal
    C}({\boldsymbol l}_\perp)=\frac{1}{{\boldsymbol  l}_\perp^{\ 2}}
   -\frac{1} {{\boldsymbol l}_\perp^{\ 2}+{ m^2_{_D}}}
\end{eqnarray}
where the two terms correspond to the exchange of a transverse and a
longitudinal gluon, respectively.  Note that the quark propagators
should be dressed in a way compatible with the resummation performed
for the vertex, in order to preserve the gauge invariance: this is the
origin of the term $-{\boldsymbol f}({\boldsymbol p}_\perp)$ under the
integral in eq.~(\ref{eq:integ-f}), which has the effect of resumming
the collisional width on the quark propagator.  From this integral
equation, it is easy to see that each extra rung in the ladder
contributes a correction of order $\alpha_s T t_{_F} \sim {\cal O}(1)$
since $t_{_F}\sim 1/g^2 T$.  Therefore, all these corrections
contribute to ${\cal O}(\alpha_s)$ to the photon rate.  Note again
that the only parameters of the QGP that enter this equation are the
quark thermal mass $m_\infty$ and the Debye screening mass $m_{_D}$.

\subsection{Some numerical results}
\begin{figure}[htbp]
\begin{flushleft}
\resizebox*{!}{6cm}{\rotatebox{-90}{\includegraphics{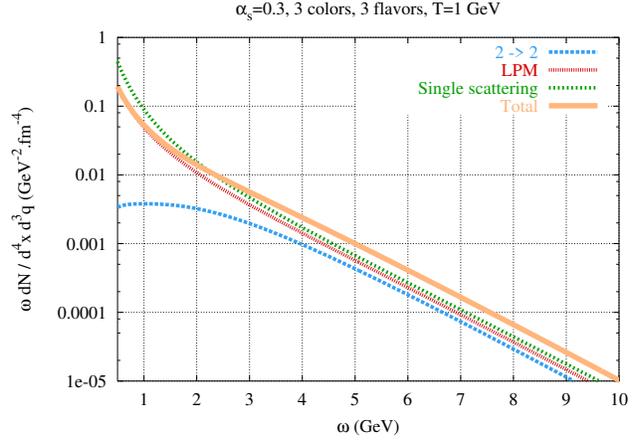}}}
\end{flushleft}
\caption{\label{fig:LPM-photon} ${\cal O}(\alpha_s)$ contributions to
    the photon production rate in a QGP. The parameters used in this
    plot are $\alpha_s=0.3$, 3 colors, 2 flavors and $T=1$~GeV. From
    \cite{ArnolMY2}.}
\end{figure}

The integral equation was solved numerically in \cite{ArnolMY2},
and the results are displayed in figure \ref{fig:LPM-photon}.
In this plot, `LPM' denotes the contribution of all the multiple
scattering diagrams, while `$2\to 2$' denotes the processes of figure
\ref{fig:processes-1}. The single scattering diagrams (figure
\ref{fig:processes-2}) are also given so that one can appreciate the
suppression due to the LPM effect (ranging typically from 15 to 30\%).

\begin{figure}[htbp]
\begin{flushleft}
\resizebox*{!}{6cm}{\rotatebox{-90}{\includegraphics{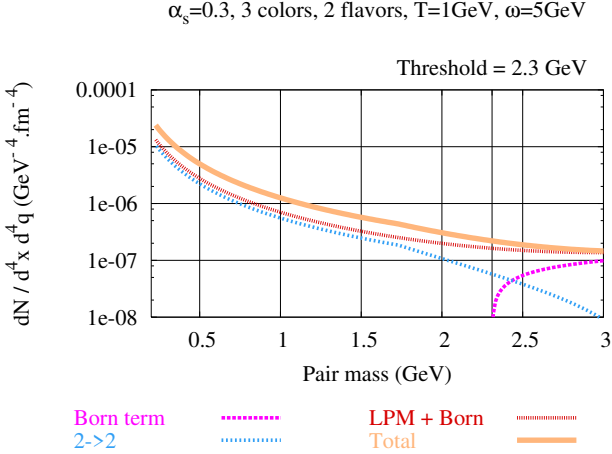}}}
\end{flushleft}
\caption{\label{fig:LPM-lepton} ${\cal O}(\alpha_s)$ contributions to
    the dilepton production rate in a QGP. From \cite{AurenGMZ1}.}
\end{figure}

Dilepton production basically suffers from the same problems, and the
solution follows the same path. Two differences are worth mentioning
here. First of all, the Drell-Yan process $q\bar{q}\to\gamma^*\to l^+
l^-$ contributes if $Q^2\ge 4 m_{\rm q}^2$. The Drell-Yan process has
been evaluated in \cite{McLerT1}, the $2\to 2$ processes have been
evaluated in \cite{AltheR1}. In addition, virtual photons have a
physical longitudinal mode that contributes to the rate of lepton
pairs. In order to take this mode into account, one must introduce a
scalar function $g({\boldsymbol p}_\perp)$ similar to ${\boldsymbol
f}({\boldsymbol p}_\perp)$, which describes the coupling of the quark
line to a longitudinal photon.  This new vertex function obeys an
integral equation \cite{AurenGMZ1} similar to eq.~(\ref{eq:integ-f}),
that resums the corrections due to multiple scatterings. This new
integral equation can also be solved numerically, and the resulting
dilepton rate (for the same parameters as in figure
\ref{fig:LPM-photon} and a total energy of the pair set to
$\omega=5~$GeV) is plotted in figure \ref{fig:LPM-lepton}.
One can see that the multiple scattering corrections are important
for all pair masses below the threshold of the Drell-Yan process.
Note also that the threshold of the tree-level process is
completely washed out when multiple rescatterings are resummed.

\section{Lattice calculations}
Attempts to calculate directly on the lattice the production
rate of dileptons in a quark-gluon plasma  appeared a few years
ago \cite{KarscLPSW1}.  What can be calculated on the lattice is the
Euclidean correlator of two vector currents, $\Pi(\tau,{\boldsymbol
x})\equiv \left<j_\mu(0,{\boldsymbol 0})j^\mu(\tau,{\boldsymbol
x})\right>$, where $\tau\in[0,1/T]$ is the Euclidean time. It is also
easy to obtain the spatial Fourier transform at zero momentum,
$\Pi(\tau,{\boldsymbol q}=0)$, by just summing over the spatial
lattice sites. The imaginary part of the real time self-energy is then
related to this object by a simple spectral representation:
\begin{equation}
{\Pi(\tau,{\boldsymbol q})}=
\int_0^{\infty} \!\!\frac{d\omega}{\pi}\, {{\rm
Im}\, \Pi_{\rm ret}{}^\mu{}_\mu(\omega,{\boldsymbol q})}\,
\frac{\cosh(\omega(\tau-1/2T))}{\sinh(\omega/2T)}\; .
\label{eq:lattice}
\end{equation}
This equation uniquely defines ${\rm Im}\, \Pi_{\rm
    ret}{}^\mu{}_\mu(\omega,{\boldsymbol q})$ if $\Pi(\tau,{\boldsymbol
    q})$ is known for all $\tau\in[0,1/T]$ and if one prescribes the
    behavior of the solution at large $\omega$.

\begin{figure}[htbp]
\begin{flushleft}
\resizebox*{!}{6cm}{\rotatebox{-90}{\includegraphics{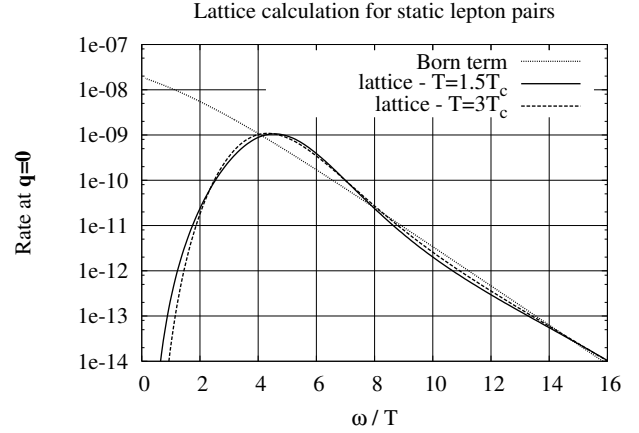}}}
\end{flushleft}
\caption{\label{fig:lattice} Lattice result for the production rate of
dileptons with ${\boldsymbol q})=0$. From \cite{KarscLPSW1}.}
\end{figure}

However, the function $\Pi(\tau,{\boldsymbol q}=0)$ is known only on
the discrete
temporal lattice sites, which prevents us  from determining uniquely ${\rm
Im}\, \Pi_{\rm ret}{}^\mu{}_\mu(\omega,0)$. This problem has been
reconsidered recently using the Maximum Entropy Method
\cite{KarscLPSW1,Asakawa:2000tr}, which is a way to take
into account prior knowledge about the solution (positivity, behavior
at the origin, etc...) in order to determine the most probable
solution compatible with the lattice data and with this a priori
information. The result obtained for zero momentum dileptons via this
method is displayed in Fig.~\ref{fig:lattice}, for two different
values of the temperature.  Note that this is a quenched lattice
simulation.
This result displays several interesting properties. At energies
above $4T$, the full rate is very close to the contribution of the
Born term, while at energies smaller than $3T$ it drops to
extremely small values. In addition, when plotted against
$\omega/T$, the curves for the two temperatures fall almost on top
of one another, indicating that the result scales like a universal
function of $\omega/T$, at least within the errors.

The suppression at small $\omega$ has attracted a lot of interest
because it contradicts expectations based on perturbation theory: the
resummation of thermal masses would indeed produce a drop of the Born
term because of threshold effects (see Eq.~(\ref{threshold})), but
higher order processes that do not have a threshold would fill the
spectrum at small $\omega$. Also, a threshold related to the quark
masses would occur at much smaller $\omega$ than $3T$, since thermal
masses are typically $m\sim gT$. Finally one may question whether the
accuracy of  lattice calculations in  the small $\omega$ regime is
not spoiled by finite volume effects.

On the other hand the polarization tensor at small frequency is
related to the electric conductivity \cite{Gupta:2003zh} by the relation:
\begin{eqnarray}
\sigma_{\rm el}=\lim_{\omega\to 0} {{{{\rm Im}\,\Pi_{\rm
        ret}}^i_i(\omega,{0})}/{6\omega}}.
\end{eqnarray}
 From this relation, one expects ${\rm Im}\,\Pi_{\rm
    ret}{}^i_i(\omega,{\boldsymbol q}={0})\propto \omega $ when
$\omega\to 0$. This implies that the static dilepton rate should
diverge when $\omega\to 0$.  Unless the electric conductivity in
quenched QCD is nearly zero for some reason, the lattice dilepton rate
disagrees with this prediction at small $\omega$.  Note that `small'
in these considerations means an $\omega$ small enough to be in the
hydrodynamical regime, i.e. $\omega \lesssim g^4 T$. In a strong
coupling theory, this regime could start as early as $\omega\sim T$.

Note that the previous argument rests on the possibility to replace
${\rm Im}\,\Pi_{\rm ret}{}^\mu{}_\mu(\omega,0)$ (involved in the
calculation of the dilepton rate) by ${\rm Im}\,\Pi_{\rm
    ret}{}^i{}_i(\omega,0)$ (which is the quantity needed to calculate
the conductivity). This is guaranteed by the Ward identity $q_\mu
\,\Pi_{\rm ret}{}^\mu{}_\mu(\omega,0)=0$. From this it follows indeed that,
unless singularities occur, $\Pi_{\rm ret}{}^{00}(\omega,0)=0$.

The electric conductivity has been calculated on the lattice by
S. Gupta \cite{Gupta:2003zh}, and a finite result was obtained. This
calculation provides an illustration of the sensitivity of the maximum
entropy method used to reconstruct the spectral functions to the prior
information. Gupta's calculation assumes explicitly that the spectral
function that he wants to determine behaves linearly in $\omega$ at
small $\omega$. The maximum entropy procedure yields then a finite
value for the slope, i.e., a finite value for the electric
conductivity. In \cite{KarscLPSW1} on the other hand, no particular
assumption is made about the behavior of the spectral function at
small $\omega$. For further discussion on the difficulty of
extracting transport coefficients from Euclidean lattice correlators,
see \cite{Aarts:2002vx,Aarts:2002cc}

\section{Discussion and outlook}
As of now, there are in fact arguments indicating that both the
perturbative calculations and the lattice calculation are incorrect at
small $\omega$. If one evaluates eq.~(\ref{eq:lattice}) at
$\tau=1/2T$, one gets a sum rule:
\begin{eqnarray}\label{eqn13}
\int_0^\infty {d\omega}\,
\frac{{{\rm Im}\,\Pi_{\rm
      ret}{}^\mu{}_\mu(\omega,{\boldsymbol q}=0)}}{{\sinh(\omega/2T)}}=
\Pi(1/2T,0) < \infty\; ,\nonumber \\
\end{eqnarray}
This sum rule is violated by all the existing analytical weak coupling
calculations (they give an infinite result). For instance, the expression in
eq.~(\ref{eqn6}) of the imaginary part of $\Pi$ diverges as $1/\omega$
at small $\omega$. The LPM effect would reduce the divergence to one
in $1/\sqrt\omega$. But the expected linear behavior is not achieved
in present approximations. In fact, none of the existing calculations  includes
correctly the dissipative effects that appear when one enters the
hydrodynamical regime ($\omega\to 0$).

As for the lattice, consider again eq.~(\ref{eqn13}). If one assumes
that the integral is dominated by the behavior of ${\rm Im}\,\Pi$ at
small $\omega$, i.e. ${\rm Im}\,\Pi^i_i(\omega, {\boldsymbol q}=0)\sim
6\sigma_{\rm el} \omega$, one obtains an estimate of the integral
\begin{eqnarray}\label{sigmabound}
    \Pi_i^i(\tau={1}/{2T},{\boldsymbol q}=0)\approx \int_0^\infty
\frac{d\omega}{\pi} \frac{\omega}{\sinh(\omega/2T)}=
    6\sigma_{\rm el}\pi T^2\; .\nonumber\\
\end{eqnarray}
One could perhaps argue that this estimate provides in fact an upper
bound for $\Pi_i^i(\tau={1}/{2T},{\boldsymbol q}=0)$ (if we admit
that the actual function never exceeds the linear extrapolation). In
order to make contact with lattice estimates, we define
$C_{_{EM}}=4\pi \alpha \sum_f e_f^2$. Then we can rewrite the equation
above as
\begin{eqnarray}
\frac{\sigma_{\rm el}}{C_{EM} T}\approx
\frac{1}{6\pi}\frac{\Pi_i^i(\tau={1}/{2T},0)}{C_{_{EM}} T^3}\; .
\end{eqnarray}
The left hand side can be obtained from Gupta's calculation and is a
number of order $7$. The right hand side is given in \cite{KarscLPSW1}, and is
weakly dependent of the temperature; it is a number of order 0.12.
These simple estimates suggest that the lattice calculations are not
fully consistent. If we would admit that eq.~(\ref{sigmabound})
provides a lower bound for the electric conductivity, then the
calculation in  \cite{KarscLPSW1} should yield a finite
$\sigma_{el}$, which is not compatible with fig.~\ref{fig:lattice}.
It is also somewhat puzzling that the value of $\sigma$ obtained in
\cite{Gupta:2003zh} is so much larger than the simple estimate based
on eq.~(\ref{sigmabound}); it would be interesting to know whether
the values of $\Pi(1/2T,0) $ obtained in
\cite{Gupta:2003zh} agree with those given in \cite{KarscLPSW1}.


\bibliographystyle{unsrt}

\begin{thebibliography}{99}

\bibitem{Sriva1}
{D.K. Srivastava}, Eur. Phys. J. {\bf C} {\bf 10}, 487 ({1999}).

\bibitem{SrivaS1}
{D.K. Srivastava, B. Sinha}, Phys. Rev. {\bf C} {\bf 64}, 034902
({2001}).

\bibitem{HuoviRR1}
{P. Huovinen, P.V. Ruuskanen, S.S. Rasanen}, Phys. Lett. {\bf B}
{\bf 535}, 109
    ({2002}).

\bibitem{AlamSRHS1}
{J. Alam, S. Sarkar, P. Roy, T. Hatsuda, B. Sinha}, Annals Phys.
{\bf 286}, 159
    ({2001}).

\bibitem{GELIS-Quark Matter}
{F. Gelis}, Nucl. Phys. {\bf A 715}, 329 (2003).
\bibitem{Weldo3}
{H.A. Weldon}, Phys. Rev. {\bf D} {\bf 28}, 2007 ({1983}).

\bibitem{GaleK1}
{C. Gale, J.I. Kapusta}, Nucl. Phys. {\bf B} {\bf 357}, 65
({1991}).

\bibitem{Gelis3}
{F. Gelis}, Nucl. Phys. {\bf B} {\bf 508}, 483 ({1997}).

\bibitem{McLerT1}
{L. McLerran, T. Toimela}, Phys. Rev. {\bf D} {\bf 31}, 545
({1985}).

\bibitem{BaierPS2}
{R. Baier, B. Pire, D. Schiff}, Phys. Rev. {\bf D} {\bf 38}, 2814
({1988}).

\bibitem{AltheAB1}
{T. Altherr, P. Aurenche, T. Becherrawy}, Nucl. Phys. {\bf B} {\bf
315}, 436
    ({1989}).

\bibitem{AltheR1}
{T. Altherr, P.V. Ruuskanen}, Nucl. Phys. {\bf B} {\bf 380}, 377
({1992}).


\bibitem{Kinoshita}
{T. Kinoshita}, J. Math. Phys. {\bf 3}, 650 (1962).

\bibitem{LeeN1}
{T.D. Lee, M. Nauenberg}, Phys. Rev. {\bf 133}, 1549 ({1964}).



\bibitem{Blaizot:2001nr}
    J.~P.~Blaizot, E.~Iancu,
    Phys.\ Rept.\  {\bf 359}, 355 (2002).


\bibitem{BraatP1}
{E. Braaten, R.D. Pisarski}, Nucl. Phys. {\bf B} {\bf 337}, 569
({1990}).

\bibitem{FrenkT1}
{J. Frenkel, J.C. Taylor}, Nucl. Phys. {\bf B} {\bf 334}, 199
({1990}).

\bibitem{Baym:1990uj}
    G.~Baym, H.~Monien, C.~J.~Pethick, D.~G.~Ravenhall,
    Phys.\ Rev.\ Lett.\  {\bf 64}, 1867 (1990).


\bibitem{KapusLS1}
{J.I. Kapusta, P. Lichard, D. Seibert}, Phys. Rev. {\bf D} {\bf
44}, 2774
    ({1991}).

\bibitem{BaierNNR1}
{R. Baier, H. Nakkagawa, A. Niegawa, K. Redlich}, Z. Phys. {\bf C}
{\bf 53},
    433 ({1992}).

\bibitem{AurenGKP1}
{P. Aurenche, F. Gelis, R. Kobes, E. Petitgirard}, Phys. Rev. {\bf
D} {\bf 54},
    5274 ({1996}).

\bibitem{AurenGKP2}
{P. Aurenche, F. Gelis, R. Kobes, E. Petitgirard}, Z. Phys. {\bf
C} {\bf 75},
    315 ({1997}).

\bibitem{AurenGKZ1}
{P. Aurenche, F. Gelis, R. Kobes, H. Zaraket}, Phys. Rev {\bf D}
{\bf 58},
    085003 ({1998}).

\bibitem{Mohan2}
{A.K. Mohanty}, Communication at the International Symposium in
Nuclear
    Physics, December 18-22, 2000, Mumbai, India.

\bibitem{SteffT1}
{F.D. Steffen, M.H. Thoma}, Phys. Lett. {\bf B} {\bf 510}, 98
({2001}).

\bibitem{AurenGZ4}
{P. Aurenche, F. Gelis, H. Zaraket}, JHEP {\bf 0205}, 043
({2002}).

\bibitem{AurenGZ3}
{P. Aurenche, F. Gelis, H. Zaraket}, JHEP {\bf 0207}, 063
({2002}).

\bibitem{Aurenche:2003ac}
   P.~Aurenche, M.~E.~Carrington, N.~Marchal,
   Phys.\ Rev.\ D {\bf 68}, 056001 (2003).

\bibitem{AurenGZ2}
{P. Aurenche, F. Gelis, H. Zaraket}, Phys. Rev. {\bf D} {\bf 62},
096012
    ({2000}).

\bibitem{LandaP1}
{L.D. Landau, I.Ya. Pomeranchuk}, Dokl. Akad. Nauk. SSR {\bf 92},
535 ({1953}).

\bibitem{LandaP2}
{L.D. Landau, I.Ya. Pomeranchuk}, Dokl. Akad. Nauk. SSR {\bf 92},
735 ({1953}).

\bibitem{Migda1}
{A.B. Migdal}, Phys. Rev. {\bf 103}, 1811 ({1956}).

\bibitem{Blaizot:1999xk}
    J.~P.~Blaizot, E.~Iancu,
    Nucl.\ Phys.\ B {\bf 557}, 183 (1999).



\bibitem{ArnolMY1}
{P. Arnold, G.D. Moore, L.G. Yaffe}, JHEP {\bf 0111}, 057
({2001}).

\bibitem{ArnolMY2}
{P. Arnold, G.D. Moore, L.G. Yaffe}, JHEP {\bf 0112}, 009
({2001}).

\bibitem{ArnolMY3}
{P. Arnold, G.D. Moore, L.G. Yaffe}, JHEP {\bf 0206}, 030
({2002}).

\bibitem{AurenGMZ1} {P. Aurenche, F. Gelis, G.D. Moore, H. Zaraket},
JHEP {\bf 0212}, 006 (2002).


\bibitem{KarscLPSW1}
{F. Karsch, E. Laermann, P. Petreczky, S. Stickan, I. Wetzorke},
Phys. Lett.
    {\bf B} {\bf 530}, 147 ({2002}).

\bibitem{Asakawa:2000tr}
    M.~Asakawa, T.~Hatsuda, Y.~Nakahara,
    Prog.\ Part.\ Nucl.\ Phys.\  {\bf 46}, 459 (2001).


\bibitem{Gupta:2003zh}
    S.~Gupta,
    Phys.\ Lett.\ B {\bf 597}, 57
(2004).

\bibitem{Aarts:2002vx}
   G.~Aarts, J.~M.~Martinez Resco,
   Nucl.\ Phys.\ Proc.\ Suppl.\  {\bf 119}, 505 (2003).

\bibitem{Aarts:2002cc}
   G.~Aarts, J.~M.~Martinez Resco,
   JHEP {\bf 0204}, 053 (2002).


\end{thebibliography}

\end{document}